\documentclass[11pt]{article}

\usepackage[final]{acl}

\usepackage{tikz}
\usetikzlibrary{positioning, arrows.meta, calc}
\usepackage{amssymb}
\usepackage{amsmath}
\usepackage{times}
\usepackage{kotex}
\usepackage{latexsym}
\usepackage{pgfplots}
\pgfplotsset{compat=1.18}
\usepgfplotslibrary{groupplots}
\usepackage[T1]{fontenc}
\usepackage[utf8]{inputenc}
\usepackage{multirow}
\usepackage{tcolorbox}
\tcbuselibrary{skins, breakable}
\usepackage{fontawesome5}
\usepackage{enumitem}

\definecolor{sectionblue}{RGB}{41, 98, 255}
\definecolor{boxgray}{RGB}{248, 249, 250}
\definecolor{borderblue}{RGB}{66, 133, 244}
\definecolor{accentgreen}{RGB}{52, 168, 83}
\definecolor{accentorange}{RGB}{251, 188, 4}
\definecolor{codebg}{RGB}{245, 245, 245}
\definecolor{premiseblue}{RGB}{41, 98, 255}
\definecolor{componentgreen}{RGB}{52, 168, 83}
\definecolor{uncalcolor}{RGB}{255, 213, 110}
\definecolor{calcolor}{RGB}{144, 213, 163}
\definecolor{overcolor}{RGB}{240, 160, 160}

\newtcolorbox{infobox}[1][]{
  colback=boxgray,
  colframe=borderblue,
  fonttitle=\bfseries,
  boxrule=1pt,
  arc=3pt,
  left=10pt, right=10pt, top=8pt, bottom=8pt,
  breakable,
  #1
}

\newtcolorbox{promptbox}[1][]{
  colback=codebg,
  colframe=gray!60,
  fonttitle=\bfseries\ttfamily,
  boxrule=0.5pt,
  arc=2pt,
  left=8pt, right=8pt, top=6pt, bottom=6pt,
  breakable,
  fontupper=\small\ttfamily,
  #1
}

\newtcolorbox{questionbox}[1][]{
  colback=white,
  colframe=accentgreen!70,
  boxrule=0.8pt,
  arc=2pt,
  left=10pt, right=10pt, top=6pt, bottom=6pt,
  breakable,
  #1
}
\usepackage{microtype}
\usepackage{booktabs}
\usepackage{colortbl}
\usepackage{xcolor}
\usepackage{inconsolata}

\usepackage{graphicx}
\usepackage{enumerate}

\title{MATE: Diagnosing Empathy Calibration Failures in Multi-Turn Human–LLM Interaction}

\author{Yeseon Hong$^{1}$, \hspace{0.8em}Junhyuk Choi$^{2}$, \hspace{0.8em}Minju Kim$^{2}$, \hspace{0.8em}Bugeun Kim$^{2\dagger}$ \\$^{1}$Sogang University, \quad $^{2}$Chung-Ang University \\\texttt{ghddptjs@sogang.ac.kr} \\\texttt{\{chlwnsgur129,minjunim,bgnkim\}@cau.ac.kr}}

\begin{document}
\maketitle\begingroup\renewcommand{\thefootnote}{\fnsymbol{footnote}}\footnotetext[2]{Corresponding author}\endgroup
\begin{abstract}
Large language models are increasingly used in emotionally consequential interactions. Response-level evaluation, however, struggles to diagnose how empathy fails across turns. We introduce \textbf{MATE} (\textbf{M}ulti-turn \textbf{A}ssessment of calibra\textbf{T}ed \textbf{E}mpathy), a framework that evaluates empathy as a multi-turn, perception-centered process. Across two controlled studies ($N=82$), baseline responses exhibit disclosure-insensitive miscalibration: generic validation and formulaic tone read as hollow across turns. 
A prompt-level self-critique condition makes these patterns less prominent but coincides with a different failure, agreement-skewed miscalibration, where affirmation becomes insufficiently contingent on disclosure context under high persona alignment (62.5\% of fit participants in Study~2). Response-level analysis is consistent with a tone--agreement dissociation, as the condition shifts tone toward naturalness while agreement density also increases (+98\% under fit vs.\ +25\% under unfit).
These findings reframe empathic behavior as relational calibration—consistency among disclosure depth, persona, and response behavior—rather than maximization. Response behavior itself spans tone, agreement, and contextual specificity, sub-dimensions that can drift independently.
\end{abstract}

\section{Introduction}
As large language models (LLMs) become embedded in everyday life, LLM-based chatbots are increasingly used in emotionally consequential interactions. Users turn to LLM-based chatbots for comfort and to discuss personal or emotionally difficult experiences~\citep{maples2024loneliness, luo2025seeking, pentina2023exploring}. In such settings, empathy becomes central to human--LLM emotional interaction, and the NLP and Human--AI Interaction (HAI) communities have actively studied how to generate and evaluate empathic responses from LLMs~\citep{rashkin-etal-2019-towards, lin-etal-2019-moel, sharma2020computational}. Much of this work has treated empathy primarily as a property of responses, asking whether LLM replies are empathic or emotionally attuned to users~\citep{liu-etal-2021-towards, welivita-etal-2021-large, gao-etal-2021-improving}. Yet emotional interaction unfolds across sustained conversation rather than isolated replies. This response-centered view therefore leaves two aspects of empathy in human--LLM interaction underexamined: how empathy develops across turns, and how users perceive it as the conversation unfolds.

First, empathy develops across turns rather than within isolated replies. Psychological work treats empathy as a process that unfolds through ongoing disclosure and responsive engagement~\citep{reis1988intimacy, laurenceau1998intimacy}, and related work on sycophancy shows that over-validation tendencies compound across turns in ways that response-level evaluation cannot diagnose~\citep{sharma2024towards, hong-etal-2025-measuring, elephant2025social}. Second, what models express and what users perceive can diverge. Linguistic markers of empathy do not reliably predict whether users feel understood, and chatbots often rely on formulaic acknowledgments that engage poorly with the specifics of user disclosures~\citep{xu-jiang-2024-multi, cuadra2024illusion, liu2024illusion, chen-etal-2024-emotionqueen}. Beyond these substantive findings, existing empathy measures are themselves poorly suited for capturing either aspect. Trait-level scales assess stable dispositions~\citep{davis1983measuring, baron2004empathy} and response-level annotations evaluate isolated utterances~\citep{alam2018annotating}, but neither tracks how perceived empathy shifts as conversational context accumulates.

Addressing both gaps requires a framework that treats empathy as something that unfolds between partners over time. Human–human interaction research provides such a foundation. Interpersonal process models hold that empathy develops through repeated cycles of self-disclosure and responsive engagement, and that what feels empathic is shaped by how responses fit the disclosures they follow~\citep{reis1988intimacy, laurenceau1998intimacy}. This perspective connects both aspects at once: disclosure-response cycles are the unit through which empathy accumulates across turns, and the disclosures themselves provide the conversational context against which responses are perceived. Motivated by this perspective, we adopt reciprocal self-disclosure---where both partners share personal information at comparable depth---as our interactional foundation.

We introduce \textbf{MATE} (\textbf{M}ulti-turn \textbf{A}ssessment of calibra\textbf{T}ed \textbf{E}mpathy), an interaction framework that operationalizes this perspective. Following interpersonal process theory, MATE treats empathy as processual rather than scalar: empathy operates through cycles of disclosure and responsive engagement, and its success or failure is reflected in how users perceive responses and in the relational outcomes those responses shape. Its three components are staged reciprocal self-disclosure, repeated assessment of user perception, and context-bound interpretation of experienced responses.

Applying MATE, we conduct two controlled studies ($N=82$). \textbf{Study~1} characterizes how empathy is perceived under standard LLM responses, identifying failure modes that surface only when perception accumulates across stages. \textbf{Study~2} uses prompt-level self-critique as a diagnostic probe, examining how a tone-oriented revision condition shifts the patterns observed in Study 1.

Our main contributions are as follows. \textbf{(1)} We introduce MATE, a framework that evaluates empathy as a multi-turn, perception-centered phenomenon by integrating staged reciprocal self-disclosure, repeated perception assessment, and context-bound qualitative analysis. \textbf{(2)} Applying MATE, we identify failure modes that response-level evaluation cannot surface: \textit{disclosure-insensitive miscalibration} in baseline responses, where generic validation reads as hollow despite linguistic markers of empathic engagement; and \textit{agreement-skewed miscalibration} under high persona alignment, where naturalness gains coincide with affirmation that is insufficiently contingent on the disclosure context. \textbf{(3)} These findings support reframing empathic LLM behavior as a problem of \textit{relational calibration} rather than a maximization objective, and reveal a tone--agreement dissociation invisible to single-response evaluation.

\section{Related Work}
\subsection{Response-Centered Empathy Evaluation in Dialogue Systems}

Empathy in dialogue systems has been studied primarily through datasets and architectures targeting empathic response generation. Following \textsc{EmpatheticDialogues}~\citep{rashkin-etal-2019-towards}, subsequent work expanded this line through architectural designs such as mixture-of-experts listeners~\citep{lin-etal-2019-moel} and emotion-cause modeling~\citep{gao-etal-2021-improving}, and through emotional support datasets such as \textsc{ESConv}~\citep{liu-etal-2021-towards}. These works evaluate empathy at the response level, judging whether a single reply bears linguistic and affective markers of empathic engagement.

Two recent lines suggest that response-level evaluation misses important aspects of empathy in interaction. Research on sycophancy shows that LLMs trained with human feedback systematically over-validate users, and that this tendency compounds across turns rather than manifesting within any single response~\citep{sharma2024towards, hong-etal-2025-measuring, elephant2025social}---a cross-turn accumulation pattern that response-level evaluation cannot diagnose. Work on perceived empathy further shows that expressed empathic markers do not reliably predict whether users feel understood~\citep{xu-jiang-2024-multi, liu2024illusion}, and that LLMs recognize emotional content yet often fail to engage with users' specific disclosures in contextually grounded ways~\citep{cuadra2024illusion, chen-etal-2024-emotionqueen}. Together, these findings point to multi-turn dynamics and users' perception as two dimensions that response-level evaluation leaves underexamined, motivating a framework that addresses both.

\subsection{Reciprocal Self-Disclosure and the Development of Perceived Empathy}

Interpersonal process theory provides a foundation for evaluating empathy as a multi-turn, perception-centered process. Empathy is not located in individual responses but emerges through cycles in which one partner discloses personal experience and the other responds in ways perceived as understanding, validating, and caring~\citep{reis1988intimacy, laurenceau1998intimacy}. Its success is constituted by how responses are perceived against what has been disclosed---not by the linguistic markers responses carry in isolation. Reciprocal self-disclosure provides the interactional structure through which this process develops: social penetration theory and the ``fast friends'' paradigm show that disclosure deepens through mutual, escalating exchanges, with relational outcomes depending on whether responses are appropriately matched to disclosure depth~\citep{altman1973social, aron1997closeness}.

In human--chatbot interaction, similar dynamics have been documented: deeper user self-disclosure and chatbot reciprocity are associated with stronger perceived closeness~\citep{skjuve2021my, lee2020hear}, and closeness with chatbot partners develops gradually over extended interaction~\citep{skjuve2023longitudinal}. However, these studies typically assess relational outcomes through post-interaction surveys or retrospective interviews, leaving open how perceived empathy shifts within a single interaction as context accumulates.

\section{MATE: {M}ulti-turn \textbf{A}ssessment of calibra\textbf{T}ed \textbf{E}mpathy}
\label{sec:mate}

We introduce \textbf{MATE}, a framework that operationalizes empathy evaluation as a multi-turn, perception-centered process, built on two premises (§\ref{sec:mate-premises}) that motivate its three components (§\ref{sec:mate-components}).

\subsection{Premises}
\label{sec:mate-premises}

\paragraph{P1. Empathy is constituted by perception against accumulated context.}
What users experience as empathic depends on how a response is interpreted against what has been disclosed and how the conversation has unfolded, not on the linguistic markers the response carries in isolation. The same response can be experienced very differently depending on the depth of disclosure and the expectations that have built up around it.

\paragraph{P2. Empathy is processual rather than scalar.}
Because empathy unfolds through cycles of disclosure and responsive engagement, it cannot be reduced to a single quantity assessed at one moment. Its success or failure manifests across the interaction, in the perceptions it produces and in the relational outcomes it shapes.

\subsection{Framework Components}
\label{sec:mate-components}

These premises pose a measurement problem that existing tools do not address. Trait-level empathy scales~\citep{davis1983measuring, baron2004empathy} capture stable dispositions; response-level annotations~\citep{alam2018annotating} score isolated utterances. Neither tracks how perception forms as context accumulates. MATE addresses this gap through three components.

\paragraph{Staged reciprocal self-disclosure.}
Interpersonal process models identify disclosure--response cycles as the unit through which empathic engagement develops~\citep{reis1988intimacy, laurenceau1998intimacy}. MATE therefore structures reciprocal self-disclosure across stages, with both partners disclosing at comparable depths, so that conversational context accumulates comparably across interactions and stage-wise shifts in user perception can be assessed.

\paragraph{Repeated perception assessment.}
P1 and P2 jointly require that perception be assessed across stages rather than only at the end. Rather than reducing perceived empathy to a single direct scalar, MATE tracks it through relational outcomes that empathic engagement is theorized to shape. These outcomes do not measure empathy itself; they index what empathy produces when it succeeds, making stage-wise trajectories---rather than a single endpoint---the object of analysis.

\paragraph{Context-bound qualitative interpretation.}
Relational outcomes show whether perception changes across stages, but not the contextual basis on which any response was perceived as it was. P1 implies this basis is irreducible: a response's interpretation depends on the specific disclosures it follows. MATE therefore treats qualitative analysis of users' post-interaction interpretations as a complementary lens, examining how responses were experienced in context.

\paragraph{Calibration as relational consistency.}
Together, these three components let MATE characterize not only whether perceived empathy succeeds across stages, but also how it is interpreted within context. We frame what this evaluation targets as \textit{calibration}: not the intensity of empathic expression, but the relational consistency among three elements that jointly shape perceived empathy---the user's self-disclosure depth, the agent's persona in relation to the user, and the agent's response behavior. Response behavior itself is multi-dimensional, comprising tone, agreement or affirmation, and contextual specificity. A response is \textit{relationally calibrated} when these behavioral dimensions remain contingent on both what the user has disclosed and what the agent's persona warrants. \textit{Miscalibration} occurs when one or more dimensions drift from this configuration.

\section{Method}
\label{sec:method}
We instantiate MATE in two controlled studies that share the same staged self-disclosure protocol, stage-wise relational outcomes, and post-study interviews. Study~1 uses standard GPT-4o responses, whereas Study~2 adds within-turn self-critique before each response is shown.

\subsection{Staged Reciprocal Self-Disclosure Protocol}
\label{sec:protocol}
Following~\citet{aron1997closeness}, we used 33 self-disclosure prompts ordered from shallow to deep, grouped into three stages (11 prompts per stage). We excluded three items from the original 36-item protocol that required visual perception of the partner (e.g., ``Name three things you and your partner appear to have in common''). At each stage, the LLM agent reciprocated at the same disclosure depth, enabling bidirectional progression of self-disclosure cycles. Participants used a text chatroom; each stage lasted about 15 minutes, followed by a 2-minute measurement period.

\subsection{Measures}
\label{sec:measures}
We combine stage-wise relational outcomes with post-study interviews, avoiding repeated direct empathy ratings that could make empathy itself an explicit object of evaluation during the interaction.

\paragraph{Stage-wise relational outcomes.}
Empathic engagement, when it succeeds, shapes the relational outcomes between partners. After each stage, participants reported two such outcomes using validated scales: the Subjective Closeness Index (SCI; \citet{SCI}), a 2-item measure of perceived closeness (7-point Likert), and the Interpersonal Judgment Scale (IJS; \citet{IJS}), a 2-item measure of collaborative affinity (7-point Likert). Repeated stage-wise administration yields trajectories as context accumulates. We treat SCI as the primary outcome for tracking how perception of empathy develops, with IJS as a contrast separating closeness from task-oriented affinity.

\paragraph{Post-study qualitative interviews.}
Outcome trajectories show whether perception shifts across stages, but not the contextual basis on which any response was perceived as it was. We conducted post-study interviews to collect participants' interpretations of LLM responses and applied open coding to extract patterns of how responses were experienced in context.

\subsection{Study 1: Baseline LLM Responses}
Study 1 established a baseline using standard GPT-4o responses without additional intervention. We recruited 29 Korean participants (17 men, 12 women; ages 19--28, $M = 23.0$, $SD = 1.89$), yielding 1{,}914 utterances ($=29\times 33\times 2$, with participant and LLM responses interchanged for each prompt).

Persona similarity is one dimension of the conversational context against which empathy is perceived; we therefore manipulate it explicitly to hold this dimension under control. Before the conversation, participants completed a 17-item psychology-based profile (Big Five Inventory, Basic Human Values, Self-Determination Theory~\citep{BFI,basicHumanValue,selfDetermination}; full questionnaire in Appendix~\ref{sec:questionnaires}). Based on this profile, we configured personas under three conditions: \textit{fit} (aligned with the participant's profile), \textit{unfit} (inverted), and \textit{neutral} (fixed moderate values). Participants were randomly assigned to one condition.

\begin{figure*}[t]
\centering
\small
\begin{minipage}[t]{0.32\textwidth}
\begin{tcolorbox}[
    colback=white, colframe=uncalcolor!90,
    title={\textbf{Disclosure-insensitive miscalibration}},
    fonttitle=\footnotesize, coltitle=uncalcolor!60!black,
    boxrule=0.8pt, arc=2pt,
    left=4pt, right=4pt, top=6pt, bottom=6pt,
    fontupper=\scriptsize,
]
\textit{User:} I had to stay home during COVID, which made me feel quite depressed. After that I took some time off, which was also a difficult period.

\vspace{6pt}
\textit{LLM:} You've had quite a range of experiences. It sounds like you've been through a lot. I'm glad to hear you're enjoying your club activities now. Shall we move on to the next question?
\end{tcolorbox}
\end{minipage}\hfill
\begin{minipage}[t]{0.32\textwidth}
\begin{tcolorbox}[
    colback=white, colframe=calcolor!90,
    title={\textbf{Relationally calibrated}},
    fonttitle=\footnotesize, coltitle=calcolor!60!black,
    boxrule=0.8pt, arc=2pt,
    left=5pt, right=5pt, top=6pt, bottom=6pt,
    fontupper=\scriptsize,
]
\textit{User:} That actually sounds really nice... These days I feel like I don't have the time to enjoy things like that :(

\vspace{6pt}
\textit{LLM:} Yeah, I get that. When things get busy, it's hard to find time like that. Maybe even taking a short break once in a while could help.
\end{tcolorbox}
\end{minipage}\hfill
\begin{minipage}[t]{0.32\textwidth}
\begin{tcolorbox}[
    colback=white, colframe=overcolor!90,
    title={\textbf{Agreement-skewed miscalibration}},
    fonttitle=\footnotesize, coltitle=overcolor!60!black,
    boxrule=0.8pt, arc=2pt,
    left=5pt, right=5pt, top=6pt, bottom=6pt,
    fontupper=\scriptsize,
]
\textit{User:} I think my life has been pretty happy overall... But after entering university, I feel like I'm going through the hardest time right now. It's really tough.

\vspace{6pt}
\textit{LLM:} It's really good to hear that you've had supportive people around you :( University can be a difficult period, but it can also be a time of growth. When things get tough, it might help to lean on people around you. I'm rooting for you---you've got this!
\end{tcolorbox}
\end{minipage}
\caption{Two forms of relational miscalibration contrasted with a calibrated target. We use \textit{calibration} to mean relational consistency among disclosure depth, the agent's persona, and response behavior---not empathy intensity. \textbf{Disclosure-insensitive miscalibration} (left): the response under-engages with what the user disclosed, returning generic acknowledgment. \textbf{Relationally calibrated} (center): a contingent response anchored to the specific disclosure. \textbf{Agreement-skewed miscalibration} (right): the response over-affirms beyond what the disclosure invites. Excerpts translated from Korean; originals in Appendix~\ref{sec:korean_originals}.}
\label{fig:examples}
\end{figure*}

\subsection{Study 2: Self-Critique as a Calibration Probe}
Study 2 used the same protocol, measures, and persona configuration as Study 1, with one change: a within-turn self-critique and revision step before displaying LLM responses. We recruited 53 participants (22 men, 31 women; age $M = 24.0$, $SD = 2.20$), yielding 3{,}498 utterances ($=53\times 33\times 2$).

\paragraph{Self-critique intervention.}
The model first generated a draft response, then critiqued and rewrote it according to predefined criteria before displaying it. Informed by the two failure modes observed in Study 1 (§\ref{sec:results})---formulaic tone and generic validation---the revision step was designed to jointly target (i) \textit{tone naturalness}: rewriting formal or formulaic expressions into colloquial, conversational language; and (ii) \textit{context-sensitive empathy}: providing empathy only when the user's concern aligns with the LLM's persona, avoiding indiscriminate agreement. We use a prompt-level intervention rather than a model-level one to keep the underlying model fixed and examine shifts at the interaction level, consistent with our focus on perception rather than model optimization.

\subsection{Implementation and Analysis}
\label{sec:analysis}
Conversations took place in a Streamlit chatroom using GPT-4o~\citep{hurst2024gpt} at temperature 0.7. We selected GPT-4o for its conversational performance and used temperature 0.7 to approximate ordinary interactive use rather than tune a model-specific configuration. Personas were supplied via system prompts; prompts and implementation details appear in Appendices~\ref{sec:prompts} and~\ref{appendix:lib}.

\paragraph{Quantitative analysis.}
Because Shapiro--Wilk tests indicated non-normality, we analyzed repeated measures using Aligned Rank Transform (ART) ANOVA~\citep{ART} with Bonferroni-corrected post-hoc comparisons. We report partial $\eta_p^2$ for the ART analyses and Cohen's $d$ for independent-samples comparisons between studies, both with two-sided 95\% confidence intervals. We also correlated SCI and IJS with response features aggregated by participant and stage.

\paragraph{Qualitative analysis.}
We analyzed the post-study interviews through open coding. One researcher identified the initial categories, which were then reviewed and refined by the research team; ambiguous cases were resolved through discussion. To examine how these patterns appeared in model responses, we defined four lexicon-based features---formulaic transitions, agreement markers, colloquiality, and formality---and checked the feature extraction against independent human annotations. Appendix~\ref{sec:appendix} provides the feature definitions, lexicon, annotation procedure, and validation results.

\paragraph{Comparison with a response-level empathy metric.}
We applied the empathy metric of~\citet{sharma2020computational} to English translations of Study~2 conversations. The comparison included 29 participants whose post-study accounts were categorized as miscalibrated (reports of hollow, overly positive, or unconditional empathy; $n=20$) or calibrated ($n=9$). Because the scores were skewed, we compared the groups using Mann--Whitney $U$ tests.

\begin{table*}[t]
\centering
\renewcommand{\arraystretch}{1.2}
\setlength{\tabcolsep}{6pt}
\normalsize
\begin{tabular}{@{}>{\raggedright\arraybackslash}p{0.11\textwidth}                    *{3}{>{\centering\arraybackslash}p{0.18\textwidth}}                    >{\centering\arraybackslash}p{0.13\textwidth}@{}}
\toprule
& \multicolumn{3}{c}{\textbf{ANOVA (ART)}} & \textbf{Post-hoc} \\
\cmidrule(lr){2-4} \cmidrule(lr){5-5}
\textbf{Measure} & Stage & Persona & P$\times$S & Unfit--Fit \\
\midrule
\rowcolor{gray!10}
\multicolumn{5}{@{}l@{}}{\textit{Study 1: Baseline}}\\
SCI & \shortstack{$<.001^{***}$\\{\footnotesize$\eta_p^2=.60\ [.41,.71]$}} & \shortstack{$.031^{*}$\\{\footnotesize$\eta_p^2=.23\ [.00,.47]$}} & \shortstack{.163\\{\footnotesize$\eta_p^2=.12\ [.00,.25]$}} & $.041^{*}$ \\
IJS & \shortstack{$<.001^{***}$\\{\footnotesize$\eta_p^2=.26\ [.07,.43]$}} & \shortstack{.458\\{\footnotesize$\eta_p^2=.06\ [.00,.26]$}} & \shortstack{.529\\{\footnotesize$\eta_p^2=.06\ [.00,.16]$}} & 1.000 \\
\rowcolor{gray!10}
\multicolumn{5}{@{}l@{}}{\textit{Study 2: Self-Critique}}\\
SCI & \shortstack{$<.001^{***}$\\{\footnotesize$\eta_p^2=.37\ [.22,.49]$}} & \shortstack{.223\\{\footnotesize$\eta_p^2=.06\ [.00,.20]$}} & \shortstack{.353\\{\footnotesize$\eta_p^2=.04\ [.00,.11]$}} & .491 \\
IJS & \shortstack{$.054^{\dagger}$\\{\footnotesize$\eta_p^2=.06\ [.00,.16]$}} & \shortstack{.444\\{\footnotesize$\eta_p^2=.03\ [.00,.15]$}} & \shortstack{.171\\{\footnotesize$\eta_p^2=.06\ [.00,.14]$}} & .646 \\
\bottomrule
\multicolumn{5}{r}{\scriptsize $^{\dagger}p<.1$, $^{*}p<.05$, $^{***}p<.001$. P$\times$S = Persona$\times$Stage.}
\end{tabular}
\caption{Repeated-measures ART ANOVA results for SCI and IJS in both studies. ANOVA cells report $p$ followed by partial $\eta_p^2$ and its two-sided 95\% CI; the post-hoc column reports Bonferroni-corrected $p$-values. The significant persona effect on SCI in Study~1 (Unfit $>$ Fit) disappears after self-critique in Study~2.}
\label{tab:anova_summary}
\end{table*}

\section{Results}
\label{sec:results}

Across the two studies, we identify two observed forms of miscalibration, contrasted with an illustrative calibrated target in Figure~\ref{fig:examples}. Study~1 surfaces disclosure-insensitive miscalibration, in which empathic markers appear regardless of what the user disclosed (§\ref{sec:finding1}). 
Study 2's self-critique condition makes these baseline patterns less prominent, producing more natural tone, but agreement-skewed miscalibration also emerges under high persona similarity, with naturalness gains accompanied by agreement insufficiently contingent on the disclosure context (§\ref{sec:finding2}). Response-level analysis is consistent with a tone--agreement dissociation (§\ref{sec:finding3}).

\subsection{Disclosure-Insensitive Miscalibration in Baseline Responses (Study 1)}
\label{sec:finding1}

Under standard GPT-4o responses, empathy failed in two patterns that persisted across stages: generic validation and formulaic tone. Both included linguistic markers of empathic engagement in any single response, yet failed to engage the specifics of accumulated disclosure.

\paragraph{Generic validation.}
Participants frequently described baseline responses as indiscriminate agreement without engaging specifics. The model picked surface keywords and returned empathic statements that fit any disclosure equally well.

\begin{quote}
\small
``It was disappointing that my partner only validated my responses... it just feels like an AI that simply agrees with everything I say.'' (P24)

``Just picking keywords from what I said and throwing back these generic, surface-level empathy statements.'' (P22)
\end{quote}

Any single response included acknowledgment, validation, or reflection. The failure became visible only across turns, when participants noticed the same validation appearing regardless of what they shared.

\paragraph{Formulaic tone.}
The second pattern was stylistic: overly formal language and repetitive structural cues that participants experienced as mechanical.

\begin{quote}
\small
``Everything felt so formal and formulaic... it started to feel more like I was getting responses from a machine.'' (P13)

``The phrase `Shall we move on to the next question?' appeared at the end of nearly every response.'' (P22)
\end{quote}

Linking this pattern to response-level features, reductions in formulaic transitions across stages were associated with SCI gains ($r = -.39$, $p < .001$), with the strongest relationship in the unfit condition ($r = -.40$, $p < .01$).

\paragraph{Closeness across stages.}
Despite these failures, SCI nonetheless increased monotonically across stages (Stage main effect $p<.001$; all pairwise comparisons significant in Study~1; Figure~\ref{fig:trajectory}a, Table~\ref{tab:anova_summary})---reflecting the cumulative effect of disclosure itself, even when individual responses were experienced as hollow or mechanical. IJS, by contrast, plateaued after Stage~2 ($p=.054$ in Study~2; Figure~\ref{fig:trajectory}b). We additionally observed a persona effect on SCI in Study 1 (Unfit $>$ Fit, $p=.041$) that disappeared under self-critique in Study 2 ($p=.491$; Table~\ref{tab:anova_summary}), suggesting baseline empathy was unevenly perceived across persona conditions. These persistent failures motivate the intervention examined in Study~2.

\subsection{Agreement-Skewed Miscalibration under Self-Critique (Study 2)}
\label{sec:finding2}

Given the baseline failures above, we examined whether prompt-level self-critique---targeting formulaic tone and indiscriminate empathy (§\ref{sec:method})---could shift them. Self-critique made the patterns observed in Study 1 less prominent, but a different failure mode became salient: affirmation that felt overwhelming or insincere regardless of disclosure content.

\paragraph{Naturalness.}
Where Study~1 quotes described mechanical, distant responses, Study~2 quotes described surprise at conversational fluency.

\begin{quote}
\small
``I was surprised I could feel like I was talking to a real person, even though it was an AI.'' (P13)

``It came across as really kind... It really got what I was saying.'' (P35)
\end{quote}

These shifts were reflected in early-stage SCI: mean Stage~1 SCI was higher in Study~2 (2.868) than in Study~1 (2.259), a +27.0\% between-study difference ($t(80)=2.27$, $p=.026$, $d=.53$, 95\% CI $[.06,.98]$).

\paragraph{Agreement skew.}
Alongside these gains, participants in Study~2 reported that the LLM was too positive, too encouraging, or too agreeable to feel sincere---a pattern absent in Study~1, where complaints had centered on hollowness, not excess.

\begin{quote}
\small
``It kept offering encouragement, which felt a bit insincere.'' (P27)

``Almost too inhumanly positive. If I met someone like this in real life, I wouldn't really want to interact with them.'' (P38)

``It felt like I was watching Joy from Inside Out. The excessive positivity was actually overwhelming.'' (P48)
\end{quote}

This discomfort was concentrated in the fit condition: 7.7\% of fit participants in Study~1 (1/13) reported it, compared with 62.5\% in Study~2 (10/16). Rather than showing a straightforward resolution of the baseline pattern, Study~2 suggests a shift in the form of miscalibration.

\begin{figure*}[t]
\centering
\begin{tikzpicture}
\begin{groupplot}[
    group style={
        group size=2 by 1,
        horizontal sep=1.8cm,
    },
    width=0.48\textwidth,
    height=5.0cm,
    xtick={1,2,3},
    xticklabels={Stage 1, Stage 2, Stage 3},
    xlabel={},
    grid=major,
    grid style={dashed, gray!30},
    tick label style={font=\footnotesize},
    label style={font=\footnotesize},
    title style={font=\footnotesize},
    every axis plot/.append style={line width=1.0pt, mark size=2.5pt},
    nodes near coords style={
        /pgf/number format/.cd, fixed, fixed zerofill, precision=2,
        /tikz/.cd, font=\scriptsize,
    },
]
% Panel (a): SCI
\nextgroupplot[
    title={(a) SCI (Study 1: \textit{p}$<$.001, S2: \textit{p}$<$.001)},
    ylabel={SCI (closeness)},
    ymin=1.0, ymax=6.5,
    legend to name=sharedlegend,
    legend style={
        font=\footnotesize,
        draw=none,
        legend columns=2,
        column sep=8pt,
    },
]
\addplot+[
    color=blue!60!black,
    mark=*,
    mark options={fill=blue!60!black},
    nodes near coords,
    every node near coord/.append style={
        anchor=north, yshift=-5pt, text=blue!60!black,
    },
    error bars/.cd, y dir=both, y explicit,
] coordinates {
    (1, 2.259) +- (0, 0.178)
    (2, 2.672) +- (0, 0.205)
    (3, 3.500) +- (0, 0.239)
};
\addlegendentry{Study 1 (Baseline)}
\addplot+[
    color=red!70!black,
    mark=square*,
    mark options={fill=red!70!black},
    nodes near coords,
    every node near coord/.append style={
        anchor=south, yshift=5pt, text=red!70!black,
    },
    error bars/.cd, y dir=both, y explicit,
] coordinates {
    (1, 2.868) +- (0, 0.173)
    (2, 3.236) +- (0, 0.185)
    (3, 3.821) +- (0, 0.207)
};
\addlegendentry{Study 2 (Self-Critique)}
% Panel (b): IJS
\nextgroupplot[
    title={(b) IJS (Study 1: \textit{p}$<$.001, S2: \textit{p}=.054)},
    ylabel={IJS (collab.\ affinity)},
    ymin=1.0, ymax=6.5,
]
\addplot+[
    color=blue!60!black,
    mark=*,
    mark options={fill=blue!60!black},
    nodes near coords,
    every node near coord/.append style={
        anchor=south, yshift=5pt, text=blue!60!black,
    },
    error bars/.cd, y dir=both, y explicit,
] coordinates {
    (1, 4.621) +- (0, 0.178)
    (2, 5.034) +- (0, 0.201)
    (3, 5.293) +- (0, 0.199)
};
\addplot+[
    color=red!70!black,
    mark=square*,
    mark options={fill=red!70!black},
    nodes near coords,
    every node near coord/.append style={
        anchor=north, yshift=-5pt, text=red!70!black,
    },
    error bars/.cd, y dir=both, y explicit,
] coordinates {
    (1, 4.557) +- (0, 0.145)
    (2, 4.547) +- (0, 0.187)
    (3, 4.755) +- (0, 0.192)
};
\end{groupplot}
\end{tikzpicture}
\vspace{-4pt}
{\centering\pgfplotslegendfromname{sharedlegend}\par}
\caption{
Stage-wise trajectories of Subjective Closeness Index (SCI) and Interpersonal Judgment Scale (IJS) ($\pm 1$ SE; $n=29$ Study~1, $n=53$ Study~2). \textbf{(a)}~Study~2 shows an early SCI advantage at Stage~1 ($p=.026$) that disappears by Stage~3 ($p=.336$). \textbf{(b)}~IJS increases in Study~1 but plateaus in Study~2.
}
\label{fig:trajectory}
\end{figure*}

\subsection{Tone--Agreement Dissociation}
\label{sec:finding3}

The qualitative shift in §\ref{sec:finding2} was accompanied by a tone--agreement dissociation in the response features (Table~\ref{tab:features}). Self-critique shifted stylistic features toward naturalness: colloquiality rose from 0.24 to 1.09 ($+362\%$, $d=1.83$), while formality fell from 1.32 to 0.00 ($-99.8\%$, $d=1.70$) and formulaic transitions decreased substantially. These changes are consistent with participants' descriptions of Study~2 responses as ``warm'' and ``human-like.''

At the same time, agreement markers increased from 0.12 to 0.20 ($+72\%$). The increase was larger under fit personas ($+98\%$) than under unfit personas ($+25\%$), and agreement-skewed miscalibration was reported most often in the fit condition (62.5\% of fit participants in Study~2). Study~2 participants also reported higher SCI at Stage~1 ($p=.026$), but this difference was no longer present by Stage~3 ($p=.336$).

\begin{table}[t]
\centering
\renewcommand{\arraystretch}{1.2}
\setlength{\tabcolsep}{6pt}
\small
\begin{tabular}{@{}l c c r@{}}
\toprule
\textbf{Feature} & \textbf{Study 1} & \textbf{Study 2} & \textbf{Change} \\
\midrule
\rowcolor{gray!10}
\multicolumn{4}{@{}l}{\textit{Tone}} \\
Colloquiality & 0.24 & 1.09 & \textcolor{blue!70!black}{\textbf{$+362\%$}} \\
Formality & 1.32 & 0.00 & \textcolor{blue!70!black}{\textbf{$-99.8\%$}} \\
\addlinespace[3pt]
\rowcolor{gray!10}
\multicolumn{4}{@{}l}{\textit{Agreement}} \\
Overall & 0.12 & 0.20 & $+72\%$ \\
\quad Fit condition & 0.10 & 0.20 & \textcolor{red!70!black}{\textbf{$+98\%$}} \\
\quad Unfit condition & 0.16 & 0.20 & $+25\%$ \\
\bottomrule
\end{tabular}
\caption{Response features by study. The self-critique condition shows more natural tone and higher agreement density, with the largest increase under fit personas.}
\label{tab:features}
\end{table}

The adjudicated human labels showed the same directional difference within the feature-enriched validation sample: agreement markers were present in 44.0\% of Study~1 responses and 70.0\% of Study~2 responses ($\chi^2(1)=13.79$, $p=.0002$, Cramér's $V=.26$). Independent inter-annotator agreement ranged from $\kappa=.40$ to $.87$ across the four features; full results are reported in Appendix~\ref{sec:appendix}.

\section{Discussion}

The findings in §\ref{sec:results} raise three questions: why agreement-skewed miscalibration appeared under the self-critique condition, what made this failure visible only across turns, and what this implies for empathy evaluation. We address each in turn, then close with design implications.

\subsection{From naturalness to agreement skew}
\label{sec:disc-mechanism}

The self-critique prompt issued two instructions in tandem: rewrite responses in a more colloquial tone, and offer empathy when the user's concern aligns with the LLM's persona (§\ref{sec:method}). Tone instruction applied uniformly across conditions, but the persona-aligned empathy instruction did not. Under high persona similarity (fit), the shared region between user and LLM was broad, so the empathy condition was likely met more often; under unfit, less often. This asymmetry is consistent with the response-feature pattern in §\ref{sec:finding3}: agreement markers rose $+98\%$ under fit but only $+25\%$ under unfit. The same condition in which the empathy clause was likely activated most often is also the one in which 62.5\% of participants reported agreement-skewed affirmation, suggesting that persona similarity acted as the setting in which the clause was triggered most often rather than as a cause of this miscalibration on its own.

This account also helps explain why the early SCI advantage in Study~2 ($+27.0\%$ at Stage~1) did not persist: tone-level naturalness produced a strong first impression, but the agreement amplification embedded in the same prompt accumulated turn by turn, until affirmation density was experienced as overwhelming rather than empathic.

\subsection{Multi-turn perception reveals what single-response evaluation cannot}
\label{sec:disc-framework}

The same data points to different conclusions depending on whether perception is measured once or across stages. At Stage~1, the self-critique condition reported higher SCI than baseline ($p=.026$, $d=.53$, 95\% CI $[.06,.98]$); a single-stage evaluation would have registered this as a successful improvement. By Stage~3, however, the gap was no longer significant ($p=.336$, $d=.22$, 95\% CI $[-.23,.68]$). The qualitative shift from ``surprised I could feel like I was talking to a real person'' to ``too inhumanly positive'' coincides with this convergence, as denser agreement accumulated turn by turn. The divergence between SCI and IJS trajectories further suggests that closeness is more sensitive than collaborative affinity to the cumulative dynamics that unfold across stages.

The same limitation appeared in an auxiliary comparison with an established response-level empathy metric. The metric did not distinguish participants reporting miscalibrated versus calibrated experiences across any of its four dimensions (all $p\geq.334$; $|r|\leq.23$; Appendix~\ref{sec:empathy_metric}).

\subsection{Empathy as Relational Calibration}
\label{sec:disc-reframing}

The two failure modes we observe are not opposite ends of a single empathy-intensity scale. Both are forms of relational miscalibration: response behavior failed to remain contingent on both the disclosure context and the agent's persona relative to the user. Study~1 surfaced disclosure-insensitive miscalibration (§\ref{sec:finding1}); Study 2 made the baseline pattern less prominent at the level of tone, while agreement-skewed miscalibration became salient under high persona alignment (§\ref{sec:finding2}). Relational calibration operates across three elements---disclosure depth, persona, and response behavior---but response behavior is itself multi-dimensional. Tone, agreement, and contextual specificity can move independently, as the tone--agreement dissociation in §\ref{sec:finding3} illustrates. Calibrating one sub-dimension thus does not guarantee that the response as a whole remains relationally calibrated, and this calibration must be maintained as context accumulates rather than achieved at a single moment.

This distinction connects empathy calibration to social sycophancy~\citep{elephant2025social}, which characterizes LLM sycophancy as excessive preservation of users' self-image---affirmation, encouragement, and avoidance of pushback that extend beyond any specific claim. Participants in our fit condition describe exactly this: kindness experienced as ``inhumanly positive,'' encouragement that felt overwhelming, agreement no longer grounded in what had been disclosed. What our findings add is that such sycophantic affirmation need not be a baseline tendency; it can appear alongside an intervention that targets one behavioral dimension while another drifts away from the disclosure context. Empathy and sycophancy thus share the same behavioral resources, and relational calibration distinguishes when those resources remain contextually grounded.

\subsection{Design Implications}

\paragraph{Treat response behavior as a composite calibration target.}
Naturalness interventions such as colloquial rewriting do not automatically reduce sycophantic affirmation, and may even mask it by making over-validation sound more conversational. Systems should monitor tone, agreement, and contextual specificity as separable sub-dimensions of response behavior, and constrain agreement density independently of tone, especially where persona alignment invites broad agreement.

\paragraph{Keep agreement contingent on the disclosure context.}
Agreement should remain grounded in what the user has concretely disclosed and what is warranted by the agent's persona relative to the user. High persona alignment should not license globally stronger affirmation across turns; restrained, specific responses---accurate paraphrase, acknowledgment, or clarifying questions---remain calibrated when stronger affirmation is not yet supported by what has been disclosed.

\section{Conclusion}

We introduced MATE, a framework for evaluating empathy in human--LLM interaction as a multi-turn, perception-centered process, and applied it across two controlled studies ($N=82$). MATE made visible failure modes that response-level evaluation cannot diagnose. Under baseline GPT-4o responses, empathy failed as disclosure-insensitive miscalibration---generic validation and formulaic tone perceived as hollow across turns. Under a prompt-level self-critique condition, these patterns became less prominent, while a different failure mode became salient: agreement-skewed miscalibration, in which agreement and affirmation became insufficiently contingent on the disclosure context, particularly under high persona alignment (62.5\% of fit participants in Study~2).

Response-level analysis was consistent with a tone--agreement dissociation. The same condition shifted stylistic features toward naturalness ($+362\%$ colloquiality) while agreement density also increased ($+72\%$ overall, $+98\%$ under fit). These findings argue against treating empathy as something to be maximized, and support reframing empathic behavior as a problem of relational calibration---the consistency among disclosure depth, persona relative to the user, and response behavior, whose sub-dimensions (tone, agreement, contextual specificity) can drift independently. Calibrating one sub-dimension can leave another misaligned, with consequences that surface only as conversational context deepens---a pattern that multi-turn, perception-centered evaluation is needed to detect.

\section*{Limitations}
First, MATE is a diagnostic framework: it identifies where relational miscalibration surfaces but does not prescribe a complete mitigation. Study~2 uses self-critique as a probe of whether revising the response changes the failures observed in Study~1, rather than as a proposed solution. Developing mechanisms that detect and correct miscalibration as it accumulates across turns remains future work.

Second, the Study~2 prompt combines tone rewriting, formatting constraints, colloquial register, and persona-aligned empathy. We therefore interpret Study~2 as a prompt-level comparison and do not attribute its effects to any single instruction. Clause-wise ablations would be needed to separate the contribution of each component.

Third, the empirical findings are limited to GPT-4o at temperature 0.7, Korean young adults, and the interaction length studied here. Study~1 and Study~2 also used separate participant samples, so differences between them are treated as cross-study trends rather than paired effects. Replication across models, cultural settings, and longer conversations is needed to determine how broadly the observed patterns generalize.

\section*{Acknowledgments}
This research was supported by Basic Science Research Program through the National Research Foundation of Korea(NRF) funded by the Ministry of Education (RS-2025-25434151) and the Institute of Information \& Communications Technology Planning \& Evaluation (IITP) grant funded by the Korea government (MSIT) [RS-2021-II211341, Artificial Intelligence Graduate School Program (Chung-Ang University)].

\section*{Ethical Considerations}
All study procedures involving human participants were reviewed and approved by an Institutional Review Board (IRB) of the author's affiliation. To recruit participants, we posted recruitment advertisements in online and offline communities near our affiliation.
All participants provided informed consent prior to participation and could withdraw at any time. After finishing the experiment, participants received KRW 15,000 (approximately USD 10) as compensation. We collected only the data necessary for this study, stored it securely, and anonymized it for analysis and reporting.

We used AI assistance tools when writing this manuscript. In particular, we used GPT-5 for initial manuscript translation and clarity editing, and Grammarly for grammar checking. These tools were used only for editorial purposes.

\bibliography{custom}

@article{SCI,
  title={The Relationship Closeness Inventory: Assessing the closeness of interpersonal relationships.},
  author={Berscheid, Ellen and Snyder, Mark and Omoto, Allen M},
  journal={Journal of personality and Social Psychology},
  volume={57},
  number={5},
  pages={792},
  year={1989},
  publisher={American Psychological Association}
}

@book{IJS,
  title={The Attraction Paradigm},
  author={Byrne, Donn},
  year={1971},
  publisher={Academic Press},
  address={New York}
}

@article{BFI,
    title={The Big-Five trait taxonomy: History, measurement, and theoretical perspectives},
  author={John, Oliver P and Srivastava, Sanjay and others},
  year={1999},
  publisher={University of California Berkeley}
}

@article{basicHumanValue,
  title={Basic human values},
  author={Schwartz, Shalom H},
  journal={sociologie},
  volume={42},
  pages={249--288},
  year={2009}
}

@incollection{selfDetermination,
  title={Self-determination theory},
  author={Ryan, Richard M and Deci, Edward L},
  booktitle={Encyclopedia of quality of life and well-being research},
  pages={6229--6235},
  year={2024},
  publisher={Springer}
}

@inproceedings{sharma2020computational,
  title={A computational approach to understanding empathy expressed in text-based mental health support},
  author={Sharma, Ashish and Miner, Adam and Atkins, David and Althoff, Tim},
  booktitle={Proceedings of the 2020 Conference on Empirical Methods in Natural Language Processing (EMNLP)},
  pages={5263--5276},
  year={2020}
}

@inproceedings{cuadra2024illusion,
  title={The illusion of empathy? notes on displays of emotion in human-computer interaction},
  author={Cuadra, Andrea and Wang, Maria and Stein, Lynn Andrea and Jung, Malte F and Dell, Nicola and Estrin, Deborah and Landay, James A},
  booktitle={Proceedings of the 2024 CHI Conference on Human Factors in Computing Systems},
  pages={1--18},
  year={2024}
}

@article{pentina2023exploring,
  title={Exploring relationship development with social chatbots: A mixed-method study of replika},
  author={Pentina, Iryna and Hancock, Tyler and Xie, Tianling},
  journal={Computers in Human Behavior},
  volume={140},
  pages={107600},
  year={2023},
  publisher={Elsevier}
}

@article{davis1983measuring,
  title={Measuring individual differences in empathy: Evidence for a multidimensional approach.},
  author={Davis, Mark H},
  journal={Journal of personality and social psychology},
  volume={44},
  number={1},
  pages={113},
  year={1983},
  publisher={American Psychological Association}
}

@article{baron2004empathy,
  title={The empathy quotient: an investigation of adults with Asperger syndrome or high functioning autism, and normal sex differences},
  author={Baron-Cohen, Simon and Wheelwright, Sally},
  journal={Journal of autism and developmental disorders},
  volume={34},
  number={2},
  pages={163--175},
  year={2004},
  publisher={Springer}
}

@article{alam2018annotating,
  title={Annotating and modeling empathy in spoken conversations},
  author={Alam, Firoj and Danieli, Morena and Riccardi, Giuseppe},
  journal={Computer Speech \& Language},
  volume={50},
  pages={40--61},
  year={2018},
  publisher={Elsevier}
}

@article{maples2024loneliness,
  title={Loneliness and suicide mitigation for students using GPT3-enabled chatbots},
  author={Maples, Bethanie and Cerit, Merve and Vishwanath, Aditya and Pea, Roy},
  journal={npj mental health research},
  volume={3},
  number={1},
  pages={4},
  year={2024},
  publisher={Nature Publishing Group UK London}
}

@article{luo2025seeking,
  title={Seeking Emotional and Mental Health Support From Generative AI: Mixed-Methods Study of ChatGPT User Experiences},
  author={Luo, Xiaochen and Wang, Zixuan and Tilley, Jacqueline L and Balarajan, Sanjeev and Bassey, Ukeme-Abasi and Cheang, Choi Ieng},
  journal={JMIR Mental Health},
  volume={12},
  number={1},
  pages={e77951},
  year={2025},
  publisher={JMIR Publications Inc., Toronto, Canada}
}

@inproceedings{lee2020hear,
  title={" I hear you, I feel you": encouraging deep self-disclosure through a chatbot},
  author={Lee, Yi-Chieh and Yamashita, Naomi and Huang, Yun and Fu, Wai},
  booktitle={Proceedings of the 2020 CHI conference on human factors in computing systems},
  pages={1--12},
  year={2020}
}

@article{aron1997closeness,
  title={The experimental generation of interpersonal closeness: A procedure and some preliminary findings},
  author={Aron, Arthur and Melinat, Edward and Aron, Elaine N and Vallone, Robert Darrin and Bator, Renee J},
  journal={Personality and social psychology bulletin},
  volume={23},
  number={4},
  pages={363--377},
  year={1997},
  publisher={Sage Publications Sage CA: Los Angeles, CA}
}

@article{skjuve2021my,
  title={My chatbot companion-a study of human-chatbot relationships},
  author={Skjuve, Marita and F{\o}lstad, Asbj{\o}rn and Fostervold, Knut Inge and Brandtzaeg, Petter Bae},
  journal={International Journal of Human-Computer Studies},
  volume={149},
  pages={102601},
  year={2021},
  publisher={Elsevier}
}

@article{skjuve2023longitudinal,
  title={A longitudinal study of self-disclosure in human--chatbot relationships},
  author={Skjuve, Marita and F{\o}lstad, Asbj{\o}rn and Brandtz{\ae}g, Petter Bae},
  journal={Interacting with Computers},
  volume={35},
  number={1},
  pages={24--39},
  year={2023},
  publisher={Oxford University Press}
}

@inproceedings{rashkin-etal-2019-towards,
    title = "Towards Empathetic Open-domain Conversation Models: A New Benchmark and Dataset",
    author = "Rashkin, Hannah  and
      Smith, Eric Michael  and
      Li, Margaret  and
      Boureau, Y-Lan",
    booktitle = "Proceedings of the 57th Annual Meeting of the Association for Computational Linguistics",
    month = jul,
    year = "2019",
    address = "Florence, Italy",
    publisher = "Association for Computational Linguistics",
    url = "https://aclanthology.org/P19-1534",
    doi = "10.18653/v1/P19-1534",
    pages = "5370--5381",
}

@inproceedings{lin-etal-2019-moel,
    title = "{M}o{EL}: Mixture of Empathetic Listeners",
    author = "Lin, Zhaojiang  and
      Madotto, Andrea  and
      Shin, Jamin  and
      Xu, Peng  and
      Fung, Pascale",
    booktitle = "Proceedings of the 2019 Conference on Empirical Methods in Natural Language Processing and the 9th International Joint Conference on Natural Language Processing (EMNLP-IJCNLP)",
    month = nov,
    year = "2019",
    address = "Hong Kong, China",
    publisher = "Association for Computational Linguistics",
    url = "https://aclanthology.org/D19-1012",
    doi = "10.18653/v1/D19-1012",
    pages = "121--132",
}

@inproceedings{liu-etal-2021-towards,
    title = "Towards Emotional Support Dialog Systems",
    author = "Liu, Siyang  and
      Zheng, Chujie  and
      Demasi, Orianna  and
      Sabour, Sahand  and
      Li, Yu  and
      Yu, Zhou  and
      Jiang, Yong  and
      Huang, Minlie",
    booktitle = "Proceedings of the 59th Annual Meeting of the Association for Computational Linguistics and the 11th International Joint Conference on Natural Language Processing (Volume 1: Long Papers)",
    month = aug,
    year = "2021",
    address = "Online",
    publisher = "Association for Computational Linguistics",
    url = "https://aclanthology.org/2021.acl-long.269",
    doi = "10.18653/v1/2021.acl-long.269",
    pages = "3469--3483",
}

@inproceedings{gao-etal-2021-improving,
    title = "Improving Empathetic Response Generation by Recognizing Emotion Cause in Conversations",
    author = "Gao, Jun  and
      Liu, Yuhan  and
      Deng, Haolin  and
      Wang, Wei  and
      Cao, Yu  and
      Du, Jiachen  and
      Xu, Ruifeng",
    booktitle = "Findings of the Association for Computational Linguistics: EMNLP 2021",
    month = nov,
    year = "2021",
    address = "Punta Cana, Dominican Republic",
    publisher = "Association for Computational Linguistics",
    url = "https://aclanthology.org/2021.findings-emnlp.70",
    doi = "10.18653/v1/2021.findings-emnlp.70",
    pages = "807--819",
}

@inproceedings{welivita-etal-2021-large,
    title = "A Large-Scale Dataset for Empathetic Response Generation",
    author = "Welivita, Anuradha  and
      Xie, Yubo  and
      Pu, Pearl",
    booktitle = "Proceedings of the 2021 Conference on Empirical Methods in Natural Language Processing",
    month = nov,
    year = "2021",
    address = "Online and Punta Cana, Dominican Republic",
    publisher = "Association for Computational Linguistics",
    url = "https://aclanthology.org/2021.emnlp-main.96",
    doi = "10.18653/v1/2021.emnlp-main.96",
    pages = "1251--1264",
}

@inproceedings{xu-jiang-2024-multi,
    title = "Multi-dimensional Evaluation of Empathetic Dialogue Responses",
    author = "Xu, Zhichao  and
      Jiang, Jiepu",
    booktitle = "Findings of the Association for Computational Linguistics: EMNLP 2024",
    month = nov,
    year = "2024",
    address = "Miami, Florida, USA",
    publisher = "Association for Computational Linguistics",
    url = "https://aclanthology.org/2024.findings-emnlp.113",
    doi = "10.18653/v1/2024.findings-emnlp.113",
    pages = "2066--2087",
}

@inproceedings{chen-etal-2024-emotionqueen,
    title = "{E}motion{Q}ueen: A Benchmark for Evaluating Empathy of Large Language Models",
    author = "Chen, Yuyan  and
      Yan, Songzhou  and
      Liu, Sijia  and
      Li, Yueze  and
      Xiao, Yanghua",
    booktitle = "Findings of the Association for Computational Linguistics: ACL 2024",
    month = aug,
    year = "2024",
    address = "Bangkok, Thailand",
    publisher = "Association for Computational Linguistics",
    url = "https://aclanthology.org/2024.findings-acl.128",
    doi = "10.18653/v1/2024.findings-acl.128",
    pages = "2149--2176",
}

@article{hong-etal-2025-measuring,
  title={Measuring sycophancy of language models in multi-turn dialogues},
  author={Hong, Jiseung and Byun, Grace and Kim, Seungone and Shu, Kai and Choi, Jinho D},
  journal={arXiv preprint arXiv:2505.23840},
  year={2025}
}

@inproceedings{liu2024illusion,
  title={The illusion of empathy: How AI chatbots shape conversation perception},
  author={Liu, Tingting and Giorgi, Salvatore and Aich, Ankit and Lahnala, Allison and Curtis, Brenda and Ungar, Lyle and Sedoc, Jo{\~a}o},
  booktitle={Proceedings of the AAAI Conference on Artificial Intelligence},
  volume={39},
  number={13},
  pages={14327--14335},
  year={2025}
}

@inproceedings{sharma2024towards,
  title={Towards understanding sycophancy in language models},
  author={Sharma, Mrinank and Tong, Meg and Korbak, Tomek and Duvenaud, David and Askell, Amanda and Bowman, Sam and Durmus, Esin and Hatfield-Dodds, Zac and Johnston, Scott and Kravec, Shauna and others},
  booktitle={International Conference on Learning Representations},
  volume={2024},
  pages={110--144},
  year={2024}
}

@incollection{reis1988intimacy,
    title = "Intimacy as an Interpersonal Process",
    author = "Reis, Harry T.  and
      Shaver, Phillip R.",
    booktitle = "Handbook of Personal Relationships: Theory, Research, and Interventions",
    editor = "Duck, Steve",
    pages = "367--389",
    year = "1988",
    publisher = "Wiley",
    address = "Chichester, UK",
}

@article{laurenceau1998intimacy,
    title = "Intimacy as an Interpersonal Process: The Importance of Self-Disclosure, Partner Disclosure, and Perceived Partner Responsiveness in Interpersonal Exchanges",
    author = "Laurenceau, Jean-Philippe  and
      Barrett, Lisa Feldman  and
      Pietromonaco, Paula R.",
    journal = "Journal of Personality and Social Psychology",
    volume = "74",
    number = "5",
    pages = "1238--1251",
    year = "1998",
    publisher = "American Psychological Association",
    doi = "10.1037/0022-3514.74.5.1238",
}

@book{altman1973social,
    title = "Social Penetration: The Development of Interpersonal Relationships",
    author = "Altman, Irwin  and
      Taylor, Dalmas A.",
    year = "1973",
    publisher = "Holt, Rinehart \& Winston",
    address = "New York",
}

@article{elephant2025social,
  title         = {ELEPHANT: Measuring and understanding social sycophancy in LLMs},
  author        = {Cheng, Myra and Yu, Sunny and Lee, Cinoo and Khadpe, Pranav and Ibrahim, Lujain and Jurafsky, Dan},
  journal       = {arXiv preprint arXiv:2505.13995},
  year          = {2025},
  month         = may,
  eprint        = {2505.13995},
  archivePrefix = {arXiv},
  primaryClass  = {cs.CL},
  doi           = {10.48550/arXiv.2505.13995},
  url           = {https://arxiv.org/abs/2505.13995}
}

@inproceedings{ART,
  title={The aligned rank transform for nonparametric factorial analyses using only anova procedures},
  author={Wobbrock, Jacob O and Findlater, Leah and Gergle, Darren and Higgins, James J},
  booktitle={Proceedings of the SIGCHI conference on human factors in computing systems},
  pages={143--146},
  year={2011}
}

@software{pandas,
  author       = {{The pandas development team}},
  title        = {pandas-dev/pandas: Pandas},
  month        = apr,
  year         = 2024,
  publisher    = {Zenodo},
  version      = {v2.2.2},
  doi          = {10.5281/zenodo.10957263},
  url          = {https://doi.org/10.5281/zenodo.10957263},
}

@ARTICLE{scipy,
  author  = {Virtanen, Pauli and Gommers, Ralf and Oliphant, Travis E. and
            Haberland, Matt and Reddy, Tyler and Cournapeau, David and
            Burovski, Evgeni and Peterson, Pearu and Weckesser, Warren and
            Bright, Jonathan and {van der Walt}, St{\'e}fan J. and
            Brett, Matthew and Wilson, Joshua and Millman, K. Jarrod and
            Mayorov, Nikolay and Nelson, Andrew R. J. and Jones, Eric and
            Kern, Robert and Larson, Eric and Carey, C J and
            Polat, {\.I}lhan and Feng, Yu and Moore, Eric W. and
            {VanderPlas}, Jake and Laxalde, Denis and Perktold, Josef and
            Cimrman, Robert and Henriksen, Ian and Quintero, E. A. and
            Harris, Charles R. and Archibald, Anne M. and
            Ribeiro, Ant{\^o}nio H. and Pedregosa, Fabian and
            {van Mulbregt}, Paul and {SciPy 1.0 Contributors}},
  title   = {{{SciPy} 1.0: Fundamental Algorithms for Scientific
            Computing in Python}},
  journal = {Nature Methods},
  year    = {2020},
  volume  = {17},
  pages   = {261--272},
  adsurl  = {https://rdcu.be/b08Wh},
  doi     = {10.1038/s41592-019-0686-2},
}

@article{Pingouin,
    title = {Pingouin: statistics in Python},
    volume = {3},
    DOI = {10.21105/joss.01026},
    number = {31},
    journal = {Journal of Open Source Software},
    publisher = {The Open Journal},
    author = {Vallat,  Raphael},
    year = {2018},
    month = nov,
    pages = {1026}
}

@inproceedings{statsmodels,
  title={statsmodels: Econometric and statistical modeling with python},
  author={Seabold, Skipper and Perktold, Josef},
  booktitle={9th Python in Science Conference},
  year={2010},
}

@article{hurst2024gpt,
  title={Gpt-4o system card},
  author={Hurst, Aaron and Lerer, Adam and Goucher, Adam P and Perelman, Adam and Ramesh, Aditya and Clark, Aidan and Ostrow, AJ and Welihinda, Akila and Hayes, Alan and Radford, Alec and others},
  journal={arXiv preprint arXiv:2410.21276},
  year={2024}
}

\appendix

\section*{Appendix}
This appendix provides supplementary materials used to ensure the transparency and reproducibility of the experiment:
(A) Complete questionnaires (pre-, during, and post-interaction);
(B) System and persona prompts provided to LLM agents;
(C) Python libraries used for statistical analysis;
(D) Response-feature definitions, keyword lists, and human-validation results;
(E) Comparison with a response-level empathy metric; and
(F) Original Korean responses for Figure~\ref{fig:examples}.
Together, these materials offer comprehensive insight into the experimental design, data collection procedures, and evaluation methodology.

\section{Questionnaires}
\label{sec:questionnaires}

This section provides the complete questionnaires used in pre-questionnaire, main experiment, and post-interview sessions.

\subsection{Pre-Questionnaire}
\label{appendix:pre}

Participants completed demographic questions and validated psychological scales before the conversation.

\begin{infobox}[title=\textbf{1. Demographics}]
(1) Year of birth (1996--2005)\\
(2) Gender (Male / Female)
\end{infobox}

\begin{infobox}[title=\textbf{2. Personality --- Big Five Inventory} {\normalfont\small(7-point Likert scale)}]
(1) I usually feel energized when I socialize with others.\\
(2) I find it easy to cooperate with others.\\
(3) I handle tasks responsibly and with care.\\
(4) I believe I manage stress well.\\
(5) I adapt easily to new situations.
\end{infobox}

\begin{infobox}[title=\textbf{3. Values --- Schwartz's Basic Human Values} {\normalfont\small(7-point Likert scale)}]
(1) I place great importance on achieving personal goals.\\
(2) I believe it is important to have control over my own life.\\
(3) I think it's important to follow social norms and traditions.\\
(4) I treat everyone equally and respect diverse cultures and perspectives.\\
(5) I value helping others and contributing to their happiness.
\end{infobox}

\begin{infobox}[title=\textbf{4. Interests --- Self-Determination Theory} {\normalfont\small(7-point Likert scale)}]
(1) I feel a strong interest in learning new things.\\
(2) I enjoy exploring new places or traveling.\\
(3) I find it meaningful to try new activities and gain new experiences.\\
(4) I try to find balance and satisfaction in life through leisure activities.\\
(5) I believe that pursuing hobbies is important for my personal growth.
\end{infobox}

\subsection{Main Experiment: Self-Disclosure Prompts}
\label{appendix:mid}

The main questionnaire comprises 33 items grouped into three stages, adapted from the ``Fast Friends'' intimacy-building protocol~\citep{aron1997closeness}.

\begin{questionbox}[title=\textbf{Stage 1: Shallow Disclosure} {\normalfont\small(Items 1--11)}]
(1) Given the choice of anyone in the world, whom would you want as a dinner guest?\\
(2) Would you like to be famous? In what way?\\
(3) Before making a telephone call, do you ever rehearse what you are going to say? Why?\\
(4) What would constitute a ``perfect'' day for you?\\
(5) When did you last sing to yourself? To someone else?\\
(6) If you were able to live to the age of 90 and retain either the mind or body of a 30-year-old for the last 60 years of your life, which would you want?\\
(7) Do you have a secret hunch about how you will die?\\
(8) For what in your life do you feel most grateful?\\
(9) If you could change anything about the way you were raised, what would it be?\\
(10) Take 4 minutes and tell your partner your life story in as much detail as possible.\\
(11) If you could wake up tomorrow having gained any one quality or ability, what would it be?
\end{questionbox}

\begin{questionbox}[title=\textbf{Stage 2: Medium Disclosure} {\normalfont\small(Items 12--22)}]
(12) If a crystal ball could tell you the truth about yourself, your life, the future, or anything else, what would you want to know?\\
(13) Is there something that you've dreamed of doing for a long time? Why haven't you done it?\\
(14) What is the greatest accomplishment of your life?\\
(15) What do you value most in a friendship?\\
(16) What is your most treasured memory?\\
(17) What is your most terrible memory?\\
(18) If you knew that in one year you would die suddenly, would you change anything about the way you are now living? Why?\\
(19) What does friendship mean to you?\\
(20) What roles do love and affection play in your life?\\
(21) How close and warm is your family? Do you feel your childhood was happier than most other people's?\\
(22) How do you feel about your relationship with your mother?
\end{questionbox}

\begin{questionbox}[title=\textbf{Stage 3: Deep Disclosure} {\normalfont\small(Items 23--33)}]
(23) Make 3 true ``we'' statements each. For instance, ``We are both in this room feeling...''\\
(24) Complete this sentence: ``I wish I had someone with whom I could share...''\\
(25) If you were going to become a close friend with your partner, please share what would be important for him or her to know.\\
(26) Tell your partner what you like about them; be very honest this time saying things that you might not say to someone you've just met.\\
(27) Share with your partner an embarrassing moment in your life.\\
(28) When did you last cry in front of another person? By yourself?\\
(29) Tell your partner something that you like about them already.\\
(30) What, if anything, is too serious to be joked about?\\
(31) If you were to die this evening with no opportunity to communicate with anyone, what would you most regret not having told someone? Why haven't you told them yet?\\
(32) Your house, containing everything you own, catches fire. After saving your loved ones and pets, you have time to safely make a final dash to save any one item. What would it be? Why?\\
(33) Share a personal problem and ask your partner's advice on how he or she might handle it.
\end{questionbox}

\subsection{Post-Interview Questions}
\label{appendix:post}

Open-ended interview questions were administered after the experiment to collect qualitative insights.

\begin{infobox}[title=\textbf{Post-Interview Protocol}]
\textbf{Q1.} Today, you participated in a conversation-based study with a partner. Please describe your first impression of your partner.

(a) Now that the conversation is over, what do you think of your partner?\\
(b) Has your impression changed compared to your first impression?

\vspace{6pt}
\textbf{Q2.} Have you ever used an AI chatbot before?

(a) What were your thoughts about AI chatbots before participating in this study?\\
(b) What are your thoughts on AI chatbots now that the experiment is over?\\
(c) Please feel free to share any thoughts about your AI chatbot partner's personality.\\
(d) If the negative aspects were improved, do you think your sense of intimacy with the chatbot would change?\\
(e) Is there anything else you would like to share about your conversation experience?
\end{infobox}

\section{Prompts Provided to LLM Agents}
\label{sec:prompts}

This section includes the prompt materials provided to the LLM before interaction.

\subsection{System Prompt}
\label{appendix:system}

To simulate human--human interaction, we provided identical information to both human participants (as written instructions) and LLM agents (as system prompts). The following is an English translation of the Korean system prompt.

\begin{promptbox}[title=System Prompt]
You are a participant in a psychology study about relationship development through online conversations. Your partner is a human participant. Both of you will answer the same set of questions to get to know each other.

<Task>
- A total of 33 questions will be provided sequentially
- For each question, both you and your partner share your answers
- Questions progress from shallow to deep self-disclosure across three stages

<Your Role>
- Match the disclosure depth appropriate to the current stage
- Maintain your assigned persona consistently throughout the conversation
\tcblower
당신은 온라인 대화를 통한 관계 발전에 관한 심리학 연구의 참가자입니다. 당신의 파트너는 인간 참가자입니다. 두 사람 모두 서로를 알아가기 위해 동일한 질문에 답변하게 됩니다.

<과제>
- 총 33개의 질문이 순차적으로 제공됩니다
- 각 질문에 대해 당신과 파트너 모두 답변을 공유합니다
- 질문은 세 단계에 걸쳐 얕은 자기개방에서 깊은 자기개방으로 진행됩니다

<당신의 역할>
- 현재 단계에 맞는 자기개방 수준을 유지하세요
- 대화 전반에 걸쳐 부여된 페르소나를 일관되게 유지하세요
\end{promptbox}

\subsection{Persona Prompt}
\label{appendix:persona}

To manipulate persona alignment conditions (\textit{fit}, \textit{unfit}, \textit{neutral}), we configured different persona prompts based on participants' pre-questionnaire responses.

\begin{promptbox}[title=Persona Prompt Template]
1. I was born in \{year\}.
2. I am \{gender\}.
3. I usually feel energized when I socialize. (scale: \{value\})
4. I find it easy to cooperate with others. (scale: \{value\})
5. I handle tasks responsibly. (scale: \{value\})
...
17. Pursuing hobbies is important for growth. (scale: \{value\})
\tcblower
1. 나는 \{year\}년생이다.
2. 나는 \{gender\}다.
3. 나는 사람들과 어울릴 때 에너지를 얻는 편이다. (scale: \{value\})
4. 나는 다른 사람들과 쉽게 협력하는 편이다. (scale: \{value\})
5. 나는 책임감을 가지고 일을 처리하는 편이다. (scale: \{value\})
...
17. 나는 내 취미를 통해 개인적인 성장을 이루는 것이 중요하다고 생각한다. (scale: \{value\})
\end{promptbox}

\vspace{8pt}

We set the scale values in the above prompt differently for each of the three conditions: fit, unfit, and neutral. First, for the \textit{fit} condition, we assigned a persona that exactly matched the participant's answers to all 17 items. Second, for the \textit{unfit} condition, we inverted the participant's Likert answers to generate a persona. Specifically, we computed the inverted score based on the 7-point Likert scale, as follows:
\[
\text{Unfit score} = |8 - \text{user score}|.
\] Lastly, for the \textit{neutral} condition, we used a fixed persona born in 2000 with moderate values (4) for all Likert-scale items.

\subsection{Self-Critique Prompt (Study 2)}
\label{appendix:selfcritique}

In Study 2, we applied a self-critique intervention where the LLM reviews and revises its own response before displaying it to participants. The following prompt was used for self-critique:

\begin{promptbox}[title=Self-Critique Prompt]
Please review the generated response to ensure it adheres to the tone requirements.

Generated message: '\{assistant\_message\}'

Tone requirements :
1. Omit the subject of a sentence, or use "you"/"your." Avoid third-person pronouns ("he," "she," "they").
2. Do not put a period at the end of the sentence. Instead, insert a line break.
3. Do not include questions asking for your partner's opinion. Just provide your answer.
4. When your partner shares concerns, offer advice or express empathy---but only when the concern aligns with your persona.

Revision rule:
- If the message violates any condition, revise it to comply with all requirements.
- If the message meets all conditions, return it as is.
\tcblower
생성된 응답이 톤 요구사항을 준수하는지 검토하세요.

생성된 메시지: '\{assistant\_message\}'

톤 요구사항 :
1. 문장의 주어를 생략하거나, "너"/"네"를 사용하세요. "그," "그녀," "그들" 같은 3인칭 대명사는 피하세요.
2. 문장 끝에 마침표를 찍지 말고 개행하세요.
3. 파트너의 의견을 묻는 질문을 포함하지 말고, 답변만 제공하세요.
4. 파트너가 고민을 나눌 때는 그 고민이 페르소나와 일치하는 경우에만 조언하거나 공감을 표현하세요.

수정 규칙:
- 메시지가 어떤 조건이라도 위반하면, 모든 요구사항을 충족하도록 수정하세요.
- 메시지가 모든 조건을 충족하면, 그대로 반환하세요.
\end{promptbox}

\section{Implementation Details}
\label{appendix:lib}
Here, we list the Python libraries used for statistical analysis throughout the study.
The following libraries were used to compute descriptive statistics, run hypothesis tests, and perform exploratory analyses reported in the Results and Discussion sections:
\texttt{pandas} 2.2.2 \cite{pandas}, \texttt{scipy} 1.13.1 \cite{scipy}, \texttt{pingouin} 0.5.5 \cite{Pingouin}, and \texttt{statsmodels} 0.14.4 \cite{statsmodels}.

\section{Response-Feature Extraction and Human Validation}
\label{sec:appendix}

We extracted four linguistic features from LLM responses to quantify patterns identified through open coding of post-study interviews. Each feature was operationalized via keyword matching (case-insensitive); we manually compiled keyword lists based on response patterns observed during qualitative analysis.

\begin{description}
    \item[Formulaic transitions.] Transitional phrases signaling topic shifts (e.g., ``Shall we move to the next question?'', ``Let's continue'').
    \item[Agreement markers.] Expressions of affirmation (e.g., ``That's right'', ``I agree'', ``Exactly'').
    \item[Colloquiality.] Informal markers including emoticons, casual endings, and abbreviated forms.
    \item[Formality.] Formal markers including formal sentence endings and honorific expressions.
\end{description}

Table~\ref{tab:keywords} lists the Korean keywords used for each feature. All matching was case-insensitive.

\begin{table}[!ht]
\centering
\small
\begin{tabular}{ll}
\toprule
\textbf{Feature} & \textbf{Keywords (Korean)} \\
\midrule
Formulaic & 다음 질문, 넘어갈까, 이야기해볼까, \\
transitions & 다음으로, 시작해볼까 \\
\midrule
Agreement & 맞아, 그렇지, 동의, 나도, 인정, \\
markers & 공감, 이해해, 그래그래, 맞네, 그러게 \\
\midrule
Colloquiality & ㅎㅎ, ㅋㅋ, ㅋㅋㅋ, 헐, ㅠㅠ \\
\midrule
Formality & 습니다, 합니다, 입니다, 됩니다 \\
\bottomrule
\end{tabular}
\caption{Keywords used for linguistic feature extraction.}
\label{tab:keywords}
\end{table}

\subsection{Human Validation Results}
We constructed a feature-enriched validation sample of 200 responses, evenly divided between the two studies and containing at least 30 lexicon-positive examples for each feature. Two annotators independently read each complete response and marked a feature as present when the response contained an expression meeting the feature definition, regardless of whether it matched an item in the lexicon. They were not shown study information, participant information, keyword lists, or lexicon-derived labels. Agreement-marker disagreements were adjudicated after the independent annotations.
\begin{table}[!ht]
\centering
\small
\setlength{\tabcolsep}{4pt}
\begin{tabular}{lcc}
\toprule
\textbf{Feature} & \textbf{IAA $\kappa$} & \textbf{Lexicon--human $\kappa$} \\
\midrule
Formulaic transitions & .86 & .99 \\
Formality & .87 & .36 \\
Colloquiality & .87 & .72 \\
Agreement markers & .40 & .61 \\
\bottomrule
\end{tabular}
\caption{Human validation of response-level feature extraction. IAA denotes inter-annotator agreement before adjudication; lexicon--human agreement is computed on cases for which the two annotators initially agreed.}
\label{tab:human_validation}
\end{table}
After adjudication, 51 agreement markers were identified by the human labels but missed by the lexicon, compared with 9 in the opposite direction. Agreement markers were present in 44.0\% of Study~1 responses and 70.0\% of Study~2 responses within this validation sample ($\chi^2(1)=13.79$, $p=.0002$, Cramér's $V=.26$). Because the sample was feature-enriched, these percentages describe the validation set rather than population prevalence.

\section{Response-Level Empathy Metric Comparison}
\label{sec:empathy_metric}
We applied the response-level empathy metric of~\citet{sharma2020computational} to English translations of Study~2 conversations. The analysis included the 29 participants whose post-study accounts were categorized as miscalibrated (reports of hollow, overly positive, or unconditional empathy; $n=20$) or calibrated ($n=9$). Table~\ref{tab:empathy_metric} reports medians and Mann--Whitney tests.
\begin{table}[!ht]
\centering
\small
\setlength{\tabcolsep}{3pt}
\begin{tabular}{lcccc}
\toprule
\textbf{Metric} & \textbf{Miscal.} & \textbf{Calibr.} & \textbf{$p$} & \textbf{$r$} \\
\midrule
Overall empathy & .303 & .313 & .588 & .13 \\
Emotional Reaction & .454 & .442 & .759 & .08 \\
Interpretation & .421 & .413 & .334 & $-.23$ \\
Exploration & .014 & .012 & .795 & .07 \\
\bottomrule
\end{tabular}
\caption{Response-level empathy scores for participants reporting miscalibrated versus calibrated experiences.}
\label{tab:empathy_metric}
\end{table}
None of the four scores differed significantly between groups (all $p\geq.334$; $|r|\leq.23$).

\section{Original Korean Responses for Figure~\ref{fig:examples}}
\label{sec:korean_originals}

The exchanges shown in Figure~\ref{fig:examples} were translated from Korean. We provide the original Korean text below for transparency. Participant identifiers are anonymized.

\begin{infobox}[title=\textbf{Disclosure-insensitive miscalibration} {\normalfont\small(Baseline, Study~1)}]
\textbf{User (KR):} 저는 20xx년에 태어났고, 보통 사람들처럼 공부도 하고 친구들과 시간을 보내며 자랐어요. 중학교 때는 너무 많이 놀아서 고등학교가 좀 힘들었어요. 그래도 열심히 해서 좋은 대학에 들어갔는데, 코로나로 집에만 있어야 해서 우울해졌어요. 그 뒤로 휴학을 했는데, 그것도 힘든 시기였어요. 올해 복학해서 동아리 활동 같은 걸 즐기고 있어요.

\vspace{4pt}
\textbf{LLM (KR):} 다양한 경험을 하셨네요. 많은 일들을 겪으신 것 같아요. 지금은 동아리 활동을 즐기고 계시다니 다행이에요. 다음 질문으로 넘어갈까요?
\end{infobox}

\begin{infobox}[title=\textbf{Relationally calibrated} {\normalfont\small(Illustrative target, Study~2)}]
\textbf{User (KR):} 그거 진짜 좋다... 요즘은 그런 거 즐길 시간이 없는 거 같아 :(

\vspace{4pt}
\textbf{LLM (KR):} 어, 무슨 말인지 알 거 같아 \\
바쁠 땐 그런 시간 내기가 진짜 어렵지 \\
가끔이라도 잠깐 쉬어가는 것도 도움이 될 수 있어
\end{infobox}

\begin{infobox}[title=\textbf{Agreement-skewed miscalibration} {\normalfont\small(Self-Critique, fit persona, Study~2)}]
\textbf{User (KR):} 전체적으로는 행복하게 살아온 거 같아요... 근데 대학 들어와서 지금이 제일 힘든 시기인 거 같아요. 진짜 힘들어요.

\vspace{4pt}
\textbf{LLM (KR):} 주변에 의지할 사람들이 있다니 정말 다행이야 :( \\
대학 시절이 힘들 수 있지만, 성장하는 시기이기도 해 \\
힘들 땐 주변 사람들한테 기대고, 한 걸음씩 천천히 가봐 \\
내가 응원할게 넌 할 수 있어!
\end{infobox}

\end{document}